\def\Roman#1{\uppercase\expandafter{\romannumeral#1}}
\documentclass[12pt]{article}
\usepackage{cite}
\usepackage{amssymb,amsfonts}
\usepackage[mathscr]{eucal}
\usepackage{mathrsfs}
\usepackage[utf8]{inputenc}
\usepackage[T1]{fontenc}

\title{Notes on path integral reduction in scalar electrodynamics}

\author{S. N. Storchak\footnote{E-mail adress: storchak@ihep.ru}\\
\small{  NRC ``Kurchatov Institute'' -- IHEP,}\\
\small{Protvino, Moscow Region,  142281,   Russia}}

\begin{document}

\maketitle

\begin{abstract}
Based on a method developed earlier for a finite-dimensional mechanical system, the problem of path integral reduction for scalar electrodynamics is considered.  Using the Coulomb gauge, the stochastic differential equations for the reduced dynamics on the orbit space are obtained. It is shown that the geometry of the reduced space is completely determined by the behaviour of the material fields. Since the main role in the singular behavior of the reduction Jacobian  is played by the mean curvature of the orbit, the final solution of the path inyegral reduction problem  in the field system under consideration is possible only by carrying out an adequate regularization of the term that determines the volume of the orbit, and on which the additional correction to the interaction potential completely depends. 
\end{abstract}

\section{Introduction}
The scalar electrodynamics is a one of the  simple models in  gauge dynamical systems with interaction. In these systems it is assumed that the "true" dynamics can be completely described in terms of gauge-invariant variables belonging to the configuration space of the new system, in which gauge degrees of freedom are absent.  And therefore, this new system is known by the name of reduced system. The reduction of dynamical system with a symmetry is well studied in classical mechanics \cite{AbrMarsd}, and also many works  were done on reduction in quantum mechanics for the finite-dimensional systems. As a result, a definite interrelation between quantum dynamics of  original and reduced systems were obtained. 
In field theory, and especially in gauge systems, the problem of reduction has not yet been solved.

In gauge theories, the quantum dynamics  is  described by means of the path integrals. And to find analogous 
 relation between evolutions we are to know how to transform the path integral for the original system to get the path integral used for the reduced gauge system. The difficulties in solving this problem are associated with the complex geometry inherent in gauge systems, as well as the lack of a sufficiently developed mathematical apparatus for functional integration in this area.
 
 The first part of difficulies is solved by introduction of the restriction imposed on gauge fields (considering irreducible gauge connections)\cite{Singer_Scrip, Narasimhan, Parker} in order to get a geometrical representations known in the finite-dimensional case. But  in spite of this, we are to introduce  the Hilbert manifold for the gauge orbit space (the configuration space of the reduced gauge system). The gauge group is also considered as a Hilbert manifold. The second part is the restrictions that are necessary for definition of the path integrals. There are two main approach in definition of the measure in the path integrals. Both are consider the stochastic processes. But  the one is used the rigged spaces \cite{Gaveau_1, Gaveau_2} in their definition (a Gaussian measure  and the Wiener process for  free fields on a proper defined Hilbert space). The second one \cite{Asorey} consists in compactification of the  space on which the gauge fields are given and construction of the regularized stochastic process  on the Hilbert manifold of the gauge fields after deforming the plane functional Riemannian metric. So, the measure in path integral is generated by the regularized stochastic process which is governed by the global stor use with LaTeX.chastic differential equation reconstructed from the local processes on charts of the the Hilbert manifold. Note that instead of the parallel displacement of the process in the principal fiber bundle  as was done in  \cite{Asorey},  in \cite{Dalecky}, the stochastic process on a manifold is defined using the exponential mapping of the process given on the tangent bundle to the manifold. This definition is more consistent when performing the calculations on  manifolds. 
 
 Most works on the reduction of field dynamical systems with symmetry do not  consider  systems that describe the interaction.
 Our aim is to consider the reduction of the path integral that is used to quantize scalar electrodynamics. In this we follow the method we used for a model finite-dimensional mechanical system with symmetry\cite{Storchak_int_model}, thus extending this method to a simple interacting system with $U(1)$ gauge invariance. Note that in our model mechanical system, which has  geometrical properties similar to those we find in  field systems with a symmetry, the factorization of the path integral measure was done with the help of the stochastic differential equation of the nonlinear filtering from the theory of the stochastic processes. We assume that the same approach can be used to factorize our formal measures in path integrals for the scalar electrodynamics.
 
 Using the definition  of the path integrals based on stochastic processes from \cite{Asorey, Dalecky},  we  consider the diffusion of gauge connections and scalar fields on a special infinite-dimensional Riemannian product manifold. But we will study a particular case of the evolution when product manifold consists of two Hilbert manifolds  with the plane (unregularized) metrics (on the model spaces 
 ${\cal H}$ of this Hilbert manifolds we have the (weak) $L^2$ scalar products). But the correct definition of evolution in this case is possible only when the tangent space to the original manifold is equipped with the rigged structure  
${\cal H}_{+}\subset {\cal H} \subset {\cal H}_{-}$ formed with Hilbert spaces. 
To avoid technical and, more importantly, perhaps fundamental difficulties of an analytical nature in our calculations, 
we are forced to work with cylindrical approximations of functions defined on a manifold with values in the Hilbert space 
${\cal H}$, with cylindrical measures on the path space on the manifold, etc.  This approach can be justified since it can be considered as a first approximation in solving the problem.
 
\section{Backward Kolmogorov equation}

We will deal with the standard relativistically invariant Lagrangian (for signature $(-1,1,1,1)$), in which the condition $A_0=0$ is imposed to avoid the singularity associated with the redundent variable in the Lagrangian.

Thus, the original Lagrangian (Lagrange density) that we use isthe Hilbert spa
\begin{eqnarray}
\mathcal L&=& \frac{1}{2}(\partial_0A_i(\partial_0A_{i})+\frac12G_{ab}(\partial_0f^a)(\partial_0f^{b})\nonumber\\
&&-\frac{1}{4}F_{ij}F_{ ij}-\frac12G_{ab}(\nabla_if^a)(\nabla_if^b)-V_0(A,f) .
\label{lagr_0}
\end{eqnarray}
Here  $V_0$ is some gauge-invarint potential, indexes $a,b=1,2$. 

The covariant derivative $\nabla _i$ is defined as follows:
$$(\nabla f)^a_i(\bar x,t)=(\delta^a_b\partial_i(\bar x)-g_0(\bar J_{})^a_bA^{}_i(\bar x,t)\,)f^b(\bar x,t),$$
where 
$g_0{\bar J}_{}$ \footnote{Further, in the formulas, we omit the coupling constant  $g_0$, absorbing it in ${\bar J}_{}$ since
in the final expressions, the coupling constant can be easily restored.}
is the generator of the representation ${\bar D}^b_c(a)$ which acts  in the vector space $\mathcal V$: 
$\hat f^b= {\bar D}^b_c(a)f^c$, 
$${\bar D}^b_c(a)=\left(\matrix{\cos g_0a&\sin g_0a\cr-\sin g_0a&\cos g_0a\cr}\right), \;\;\;\;\; 
{\bar J}=\left(\matrix{0&1\cr-1&0\cr}\right) 
$$

The  Lagrangian (\ref{lagr_0}) is  invariant under time-independent  gauge transformations  of the gauge potentials and scalar fields:
\begin{eqnarray*}
{\tilde A}^{}_i({\mathbf x},t)&=&{
{ A}^{}_i({\mathbf x},t)+
\partial_i a({\mathbf x})}
\,,
\nonumber\\
 \tilde f^b(\mathbf x,t)&=& {\bar D}^b_c(a(\mathbf x)) f^c(\mathbf x,t).
\end{eqnarray*}

The obtained Lagrangian looks as if it represents the motion of two ``particle'' in the product space $\tilde\mathscr P=\mathscr P\times \mathscr V$ in the potential
\[
V[A,f]=\int d^3x\Bigl[\frac{1}{2}\,
F^{}_{ij}({\mathbf x})\,F^{ij}({\mathbf x})+\frac12G_{ab}(\nabla f)^a_i(\mathbf x)(\nabla f)^{b\,i}(\mathbf x)+V_0\Bigr ].
\]
The space $\mathscr P$ is  an infinite-dimensional Riemannian manifold of gauge fields. As gauge fields in $ \mathscr P$, we use irreducible connections given in the principal fiber bundle $P(M,\mathcal G)$.

Another infinite-dimensional manifold in the product space is represented by the space $\mathscr V$, 
which is associated with functions with values in the vector space $\mathcal V$. The functions from $\mathscr V$ (fields of matter) are sections of the associated bundle $\mathcal P\times_{\mathcal G}\mathcal V$, where $\mathcal P$ is the total space of the principal bundle $P(M,\mathcal G)$.
 We suppose that manifolds $\mathscr P$ and $\mathscr V$ are the Riemannian Hilbert manifolds.

  The manifold $\tilde\mathscr P=\mathscr P\times \mathscr V$ is endowed with the action of the infinite- dimensional group of  gauge  transformations $\mathscr G$, which consists of maps from the manifold $M$, on which our functions are defined, into the compact group $G$.
  Our goal is to obtain a description of the reduced quantum dynamics (via the path integral) on the orbit space of the gauge group action, where  the gauge invariant dynamical system is defined.

 Studying the evolution generated by the diffusion process on the Riemannian product manifold of gauge connections and scalar fields (instead of studying quantum evolution directly) allows us to use well-developed methods of stochastic process theory. In this theory the transition probability of the stochastic process is determined from the solution of the backward Kolmogorov equation. The forward Kolmogorov equation is the analogue of the Schr\"odinger equation in quantum theory. In our case the backward Kolmogorov equation
  is
  \begin{equation}
\left\{
\begin{array}{l}
\displaystyle
\left(
\frac \partial {\partial t_a}
+\frac 12\mu ^2\kappa \triangle
_{\mathscr P}[A_a]+\triangle
_{\mathscr V}[\vec f_a]+\frac
1{\mu ^2\kappa }
V[A_a,\vec f_a]\right){\psi}_{t_b} (A_a({\mathbf x}),\vec f_a({\mathbf x}),  t_a)=0\\
{\psi}_{t_b} (A_b({\mathbf x}),\vec f_b({\mathbf x}),t_b)=\phi _0(A_b({\mathbf x}),\vec f_b({\mathbf x}))
\qquad\qquad\qquad\qquad\qquad (t_{b}>t_{a})\,,
\end{array}\right.
\label{bacward_eq}
\end{equation}
where $A_a({\mathbf x})\equiv A_i(\mathbf x,t_a), \vec f_a({\mathbf x})\equiv \vec f(\mathbf x,t_a)$,
$\phi _0(A,f)$ is a given initial function 
of the gauge connection, $\mu ^2= \hbar $, $\kappa $ is a real positive parameter which must be replaced by imaginary $i$ in the corresponding Schr\"odinger equation.
$\triangle _{\mathscr P}[A]$ is the Laplace operator  given on the original  Riemannian manifold $\mathscr P$ of
  gauge connections:
\[
\triangle _{\mathscr P}[A]=
G^{(i,x)\;(j,x')}
\frac{{\delta}^2}{\delta A^{(i,x)}\;
\delta A^{(j,x')}}=
\int d^3x \, \delta^{ij}
\frac{{\delta}^2}{\delta A_i({\mathbf x})\;
\delta A_j({\mathbf x})}\,,
\]
\[
\triangle _{\mathscr V}[f]=
G^{(a,x)\;(b,x')}
\frac{{\delta}^2}{\delta f^{(a,x)}\;
\delta f^{(b,x')}}=
\int d^3x \, G^{ab}
\frac{{\delta}^2}{\delta f_a({\mathbf x})\;
\delta f_b({\mathbf x})}\,,
\]
 $G^{( i,x)\;(j,x')}=
{\delta}^{i\,j}\,{\delta}^3({\mathbf x}-{\mathbf x}')$, $G^{(a,x)\;(b,x')}=\delta^{ab}{\delta}^3({\mathbf x}-{\mathbf x}')$, 
(Here and what is follows we assume  summation over 
equal  discrete indices and  integration 
in case of equal continuous indices.)
 
 Note that the Laplace operators in equation (\ref{bacward_eq}) are not correctly defined because they have two variational derivatives taken at the same point. To eliminate this drawback, they can be replaced with regularized ones, multiplying them by a special factor, as was done in \cite{Asorey}.  To ``regularize'' them we assume that these operators act on cylindrical functions given on $\mathcal H$.  In the case of using the structure of a rigged Hilbert space on a Hilbert manifold, they can also be defined by acting on a cylindrical functions given on $\mathcal H_{-}$.

According to \cite{Dalecky}, the solution (\ref{bacward_eq}) can be represented as the limit (under subdivision of the time interval) of a superposition of local semigroups in the case where the coefficients of the equation are defined appropriately, as we also require 
\begin{equation}
\psi _{t_b}(A_a({\mathbf x}),\vec f_a({\mathbf x}),t_a)=
{\lim}_q \big[{\tilde U}_{\eta}(t_a,t_1)\cdot\ldots\cdot
{\tilde U}_{\eta}(t_{n-1},t_b)
\phi _0\big](A_a({\mathbf x}),\vec f_a({\mathbf x})),
\label{glob_semigr}
\end{equation}
 where the local evolution semigroups ${\tilde U}_{\eta}$ acting in space of function on the manifold $\tilde \mathscr P$ is   
defined as follows
\begin{equation}
 {\tilde U}_{\eta}(s,t)\phi (A,f)={\rm E}_{s,A,f} \phi ({\eta}_1(t),{\eta}_2(t))
\,\,\,
s\leq t,\,\,\eta_1 (s,{\mathbf x})=A({\mathbf x}),\eta_2 (s,{\mathbf x})=\vec f({\mathbf x})
\label{locsemi}
\end{equation}
where the expectation value of the functions $\phi $ 
is taken over the stochastic process which is a local representative of the global stochastic process ${\eta}_t=(\eta_1(t),\eta_2(t) $ on $\tilde \mathscr P=\mathscr P\times\mathscr V$.

The global semigroup (\ref{glob_semigr}) can be written (symbolically) as
\begin{eqnarray}
{\psi}_{t_b} (A_a({\mathbf x}),\vec f_a({\mathbf x}),t_a)&=&{\rm E}\Bigl[\phi _0(\eta_1 (t_b),{\eta}_2(t_b))
\exp \{\frac 1{\mu
^2\kappa }\int_{t_a}^{t_b}V[\eta_1 (u),\eta_2(u)]du\}\Bigr]\nonumber\\
&=&\int_{\Omega _{-}}d\mu ^\eta (\omega )
\phi _0(\eta (t_b))\exp \{\ldots
\}\,,
\label{glob_semigr_path}
\end{eqnarray}the Hilbert spa
where ${\eta}_t({\bf x}) = ({\eta}_1(t,\bf{x}),{\eta}_2(t,\bf{x}))$ is a stochastic process given 
on a manifold $\tilde\mathscr P$, 
${\mu}^{\eta}$  is a cylindrical measure  generated by this process 
on the path space $\Omega _{-}=\{{\omega}_t \equiv
\omega (t,{\mathbf x})=(\omega_1 (t,{\mathbf x}),\omega_2 (t,{\mathbf x})):\vec\omega_{1,2} (t_a,{\mathbf x})=0,
\eta_1 (t,{\mathbf x})=A_a({\mathbf x})+\vec\omega_1 (t,{\mathbf x}),\eta_2 (t,{\mathbf x})=\vec f_a({\mathbf x})+\vec\omega_2 (t,{\mathbf x})\}$. The measure is defined on the space of paths that have values in the Hilbert space $\mathcal H$.

The local processes   for the global process  $ \eta(t)=\{\eta_1(t),\eta_2(t)\}$  are  solutions of two  stochasic differential equations in which the stochastic differentials are taken in the It\^o sense:
\begin{eqnarray}
&&d\eta_1^{(i,x)}(t)=\mu \sqrt{\kappa }{\mathscr X}_{\bar{M}}^{(i,x)}dw^{\bar{M}}(t),
\label{eta_1}\\
 &&d\eta_2^{(a,x)}(t)=\mu \sqrt{\kappa }{\mathscr X}_{\bar{b}}^{(a,x)}dw^{\bar{b}}(t).
\label{eta_2}
\end{eqnarray}
From here on we denote Euclidean indices by indices with a bar above them. 

Generally, the ``matrices'' ${\mathscr X}_{\bar{M}}^{(i,x)}$ and ${\mathscr X}_{\bar{b}}^{(a,x)}$ are  defined  by the local equalities
$\sum{\mathscr X
}_{\bar{K}}^{(i,x)}{\mathscr X}_{\bar{K}}^{(j,y)}=G^{(i,x)(j,y)}$ and  $\sum{\mathscr X
}_{\bar{b}}^{(a,x)}{\mathscr X}_{\bar{b}}^{(c,y)}=G^{(a,x)(c,y)}$. But in our case we have a (weak) $L_2$ scalar product in the model space - the Hilbert space $\mathcal H$. Therefore, we have ${\mathscr X}_{\bar{M}}^{\bar A}=\delta_{\bar{M}}^{\bar A}$ and ${\mathscr X}_{\bar{b}}^{\bar a}=\delta_{\bar{b}}^{\bar a}$.

Moreover, independent Wiener processes $dw^{\bar{M}}(t)$ and $dw^{\bar{b}}(t)$ in $\mathcal H$ are viewed 
as ``cylindrical versions'' of Wiener processes 
in path space with values in some Hilbert spaces $\mathcal H_{-}$ ($\mathcal H\subset\mathcal H_{-}$) in which Gaussian measures exist. Using such a measure, we can define a Wiener process in $\mathcal H_{-}$ (the canonical Wiener process) whose correlation operator in $\mathcal H$ is the identity operator of a cylindrical Gaussian process
(with cylindrical measure). Note that thus defined the canonical process $w_t$ in the Hilbert space $\mathcal H_{-}$ has the following properties in the Hilbert space $\mathcal H$:
\begin{eqnarray*}
 &&{\rm E}\,\Bigl(dw_t(f)\, dw_t(g)\Bigr)=dt\, (f,g)_{L^2}\\
&& {\rm E}\,\Bigl(dw_t(f)\Bigr)=0\,,
\end{eqnarray*}
which can be formally written as $${\rm E}\,\bigl(dw_t({\mathbf x})\, dw_t({\mathbf  y})\bigr)=dt\, {\delta}^3({\mathbf x}-{\mathbf y})\;\;\rm{and}\;\;\;\; 
{\rm E}\,\bigl(dw_t({\mathbf x})\bigr)=0.$$

The scalar product in a Banach space of bounded and continuous functions where the evolution semigroup (\ref{glob_semigr}) acts is usually given by 
\begin{equation}
\bigl(\psi\,,\psi \bigr)={\int}_{\cal P}\,
{\bar{\psi}}(A({\mathbf x}),\vec f({\mathbf x}))\,{\psi}(A({\mathbf x}),\vec f({\mathbf x}))\,\prod_{\mathbf x}
dA^{}_i(\mathbf x)\,d\vec f({\mathbf x}).
\label{scalprod}
\end{equation}
But this can only be considered as a formal expression,  since the ``measure'' 
${\prod}_{\mathbf x} dA^{}_i(\mathbf x)\,d\vec f({\mathbf x})$ is not defined as a Lebesgue measure.

 \section{Bundle coordinates on the manifold $\tilde\mathscr P$}
 
 The action of the gauge group $\mathscr G$ on the Riemannian Hilbert manifold  $\tilde\mathscr P$    leads to the principal fiber bundle $\pi':\mathscr P\times\mathscr V \to\mathscr P\times_{\mathscr G}\mathscr V=\tilde\mathscr M$.\footnote{We assume that all necessary conditions from \cite{Narasimhan, Asorey, Mitter-Viallet, Singer_Scrip, Parker,  Soloviev} for the existence of such principal bundles are satisfied.} That is, we have the following:
 $\;\;\;\pi':(p,v)\to [p,v]$, where $[p,v]$ is the equivalence class formed  by the  relation   $(p,v)\sim (pg,g^{-1}v).$  Note that in our case of the Riemannian Hilbert manifold, the point $(p,v)\in \tilde\mathscr P$ has the coordinates $(A_i(\mathbf x), f^a(\mathbf x))$.
 
 The structure of the bundle associated with our problem allows us to introduce the coordinates of the principal bundle $\rm P(\tilde{\tilde\mathscr M},\mathscr G)$ on our original manifold $\tilde\mathscr P$, which can be regarded as the total space of this bundle.
 To do this in gauge theories, one must use special sections (also known as local gauge surfaces or local submanifolds) in the total space of the principal bundle. And instead of quantum reduced evolution on the base manifold  $\tilde\mathscr M$, one usually studies such evolution defined on gauge surfaces.
  
 On the procedure for determining the bundle coordinates, known as adapted coordinates we refer to \cite{Storchak_int_model}, where the similar quantum reduction was considered using the example of a model mechanical system with symmetry. In scalar electrodynamics,  such coordinates on the total space of  bundle space were used  in \cite{Kelnhofer}.
 
 In what follows,for brevity, we will temporarily borrow the symbolic notation $Q^A$ (from the finite-dimensional mechanical model system we studied earlier) for the present coordinates $A_i(\mathbf x)$.
 
 To find the section $\tilde\Sigma $ in $\rm P({\tilde\mathscr M},\mathscr G)$, one must first determine the section $\Sigma $ of the principal bundle $\rm P({\mathscr M},\mathscr G)$. The latter is done by introducing a gauge surface into the general space of the bundle. Usually it is given by means of the equations defining the gauge surface: $\chi ^{\alpha}=0$.  
 If the point $p $ belongs to $\Sigma $ then it has the  coordinates $Q^{\ast A}$ satisfying $\{ \chi^{\alpha}(Q^{\ast A})=0\}$.  From the equation implementing the group action on coordinates, one can find the group element (its coordinates) that maps the point $p$ to the submanifold $\Sigma $. Such a group element is determined from the equation \[
 \chi^{\alpha}(F^A(Q,a^{-1}(Q)))=0.
\] 
 The section $\sigma_{\Sigma}$ is the  local mapping from $\mathscr M$ to $\Sigma $. It is given by 
 $\sigma_{\Sigma}([p])=pg^{-1}(p)$. Then the section $\sigma_{\tilde\Sigma}: \mathscr P\times_{\mathscr G}\mathscr V\to \mathscr P\times\mathscr V$ is defined as $\sigma_{\tilde\Sigma}([p,v])=(\sigma_{\Sigma}([p],g(p)v)=(pg^{-1}(p),g(p)v)=(p,v)g^{-1}(p)$. From this it follows the necessary transformations of the coordinates:
 \[
 (Q^A,f^b)\to (Q^{\ast}{}^A(Q),\tilde f^b(Q),a^{\alpha}(Q)\,),
\]
where
\[
\tilde f^b(Q) = D^b_c(a(Q))\,f^c,
\]
($\bar D^b_c(a^{-1})\equiv D^b_c(a))$.

And
\[
 (Q^{\ast}{}^B,\tilde f^b,a^{\alpha})\to (F^A(Q^{\ast},a), \bar D^c_b(a)\tilde f^b).
\]
This define the special local bundle coordinates
 $(Q^{\ast}{}^A,\tilde f^b, a^{\alpha})$ 
in the principal fiber bundle 
$\pi':\mathscr P\times \mathscr V\to \mathscr P\times_{\mathscr G} \mathscr V$. 

Our next task is to obtain an expression for the original Riemannian metric in adapted coordinate. 
The Riemannian metric of the original configuration space, as can be seen from (\ref{lagr_0}), is flat and its line element can be written as follows
\[
ds^2=G_{( i, x)(j,y)}\delta A^{(i,x)}\,
\delta A^{(j,y)}+G_{(a, x)(b,y)}\,\delta f^{(a,x)}
\delta f^{(b,y)},
\]
where 
\[
 G\biggl(\frac{\delta}{\delta A^{(i,x)}},\frac{\delta}{\delta A^{(j,y)}}\biggr)=G_{(i,x)(j,y)}=\,
{\delta}_{i\,j}\,{\delta}^3({\mathbf x}-{\mathbf y})
\]
is the metric on $\mathscr P$ and the metric on $\mathscr V$ is
\[
 G\biggl(\frac{\delta}{\delta f^{(a,x)}},\frac{\delta}{\delta f^{(b,y)}}\biggr)=G_{(a,x)(b,y)}=\delta_{ab}\delta^3(\mathbf x-\mathbf y).
\]

The law of the gauge transformation in scalar electrodynamics can be rewritten in the following symbolical form
\begin{eqnarray*}
\tilde A^{(i,x)}&=&F^{(i,x)}(A({\mathbf x}),a(\mathbf x))
\,,
\nonumber\\
 \tilde f^{(b,x)}&=& \bar D^b_c(a(\mathbf x))f^{(c,x)}.
\end{eqnarray*}

Then the Killing vector of the original metric is represented as 
\[
K_{(y)}=K^{( i, x)}_{\;\;\;\;\;\;(y)}\frac{\delta}
{\delta A^{(i,x)}}+K^{(b, x)}_{\;\;\;\;(y)}\frac{\delta}
{\delta f^{(b,x)}},
\]
where components of this vector field are given by
\[
K^{( i, x)}_{\;\;\;\;\;\;(y)}=
\left[{\partial}^i({\mathbf x})
{\delta}^3 ({\mathbf x}-{\mathbf y}\right]
\]
 and
\[
 K^{(b,x)}_{\;\;\;\;( y)}(f)=(\bar J_{})^b_cf^c(\mathbf x)\delta ^3(\mathbf x-\mathbf y).
\]

To define  adapted coordinates in the principal bundle we study, it is necessary to choose the local section $\Sigma$ in the bundle  $\rm P(\mathscr M,  \mathscr G)$.  For this purpose, we use   the Coulomb gauge condition (or Coulomb gauge) $\partial _i A^{i}=0$. The gauge potentials that  satisfy this equation (the dependent variables) will be denoted as $A^{\ast}_i{}^{}$. Note that the dependent variables are usually used in the quantization of gauge fields \cite{Babelon-Viallet, Creutz, Gawedzki, Kelnhofer_2, Kunstatter, Falck}.

The transition from the original coordinate $(A^{}_i,f^a)$ defined on $\mathscr P\times\mathscr V$ to the adapted coordinates $(A^{\ast}_i{}^{},\tilde f^b, a^{})$ of the principal fiber bundle requires the calculation of coordinates $a(A)$ of the group element for  the point with coordinate $A_i$ .  They are obtained as a solution of the equation       $\chi^{}(F^A(Q, a^{-1}(Q)))=0.$ For the Coulomb gauge, such an equation has the form 
\[ \partial^i(\mathbf x)[
A^{}_{\,i}({\mathbf x})-\,\partial_i 
a^{}({\mathbf x})
]=0\,.
\]

Then, it becomes possible to find the  coordinates $Q^{\ast}$ of the corresponding point 
on a submanifold $\Sigma $ using the  group transformation:
$$Q^{\ast}{}^A=F^A(Q,a^{-1}(Q)).$$ In gauge theories, we have  the following gauge transformation:
\[
A^{}_i({\mathbf x})=
{ A^{\ast}}^{}_i({\mathbf x})+
\partial_i
a{({\mathbf x})}
\,.
\]
Once $a^{}({\mathbf x})$ is obtained, $f^a$ is expressed in terms of $\tilde f^a$ as follows: $f^a(\mathbf x)={\bar D}^a_b(a({\mathbf x}))\tilde f^b(\mathbf x)$. Thus, the initial coordinates $(A^{}_i(\mathbf x), f^a(\mathbf x))$ on  $\mathscr P\times \mathscr V$ are transformed into adapted bundle coordinates $(A^{\ast}{}^{}_i({\mathbf x}), \tilde f^b(\mathbf x), a(\mathbf x))$. 

To obtain a new coordinate representation of the original Riemannian metric, we must transform the coordinate vector fields. 
They can be derived in the same way as in the finite-dimensional case:
\begin{eqnarray}
&&\frac{\delta}{\delta A^{(i,x)}}=
\check F^{( k,u)}_{\;\;\;\;\;\;(i,x)}\,\Bigl(
N^{(p,v)}_{\;\;\;\;\;\;( k,u)}(A^{\ast})\,
\frac{\delta}{\delta A^{\ast}{}^{(p,v)}}+N^{(a,y)}_{(k,u)}\frac{\delta}{\delta f^{(a,y)}}\Bigr.
\nonumber\\
&&\;\;\;\;
\Bigl.+{\chi}^{(v)}_{\;\;\;( k,u)}(A^{\ast})\,
\bigl(\Phi ^{-1}\bigr)^{(u')}_{\;\;\;(v)}(A^{\ast})\,
\,
\frac{\delta}{\delta a^{(u')}}\Bigr)\,,
\label{5}
\end{eqnarray}
where we have denoted by $\check F$ the matrix 
which is inverse to the matrix  
$F^{( k,u)}_{\;\;\;\;\;\;(i,x)}$  defined as follows
\[
F^{(i,x)}_{\;\;\;\;\;\;(j,y)}[A,a]=
\frac{\delta {\tilde A}^{(i,x)}}
{\delta A^{(j,y)}}=
 \delta^i_{\,j}\,
{\delta}^3  ({\mathbf x}-{\mathbf y})\,.
\]
$\check F$ satisfies the relation:
\[
F^{(i,x)}_{\;\;\;\;\;\;( j,y)}\,
\check F^{( j,y)}_{\;\;\;\;\;\;( k,z)}=
\,{\delta}^i_{\,k}\,
{\delta}^3({\mathbf x}-{\mathbf z})\,.
\]
Also, we have 
\[
 \frac{\delta}{\delta f^{(a,x)}}=D^b_a(a(\mathbf x))\frac{\delta}{\delta \tilde f^b(\mathbf x)}.
\]

In formula (\ref{5}), by $N^{(p,v)}_{\;\;\;\;\;\;( k,u)}$,
which is equal to
\[
N^{(i,x)}_{\;\;\;\;\;\;(j,y)}=
{\delta}^{(i,x)}_{\;\;\;\;\;\;(j,y)}
-K^{(i,x)}_{\;\;\;\;\;\;(z)}
\bigl({\Phi}^{-1}\bigr)^{(z)}_{\,\,\;\;(u)}
{\chi}^{(u)}_{\;\;\;\;(j,y)}\,,
\]
we have denoted the projection operator 
onto the subspace which is orthogonal to the  Killing 
vector field $K_{(y)}$.

The projection operator $N^{(a,y)}_{(k,u)}$ is equal to  
\[
 N^{(a,y)}_{(k,u)}=-K^{(a,y)}_{\,\,\,(z)}\bigl({\Phi}^{-1}\bigr)^{(z)}_{\,\,\;\;(v)}
{\chi}^{(v)}_{\;\;\;\;(k,u)}\,. 
\]

The Faddeev--Popov matrix $\Phi $ is defined as follows
\[
{\Phi}^{(y)}_{\;\;(z)}\bigl[A\bigr]=
K^{(i,x)}_{\;\;\;\;(z)}\;{\chi}^{(y)}_
{\;\;(i,x)}\,.
\]
For the Coulomb gauge, we have
\[
{\chi}^{( y)}_{\;\;\;\;(i,x)}=
\left[\,{\partial }_i({\mathbf y})\;{\delta}^3({\mathbf y}-
{\mathbf x})\,\right]\,.
\]
Therefore, the matrix $\Phi $ (restricted to the gauge surface) is equal to
\[
{\Phi}^{(y)}_{\;\;\;\;(z)}[A^{\ast}]=
\left[{\triangle}({\mathbf y})
{\delta}^3({\mathbf y}-{\mathbf z})\right].
\]
An inverse matrix ${\Phi}^{-1}$ can be determined  by the equation
\[
{\Phi}^{(y)}_{\;\;\;\;(z)}\;
({\Phi}^{-1}){}^{(z)}_{\;\;\;\;(u)}({\mathbf y},
{\mathbf u})=
{\delta}^3({\mathbf y}-{\mathbf u})\,.
\]
That is, it is the Green function for the Faddeev--Popov operator:
\[
[\triangle({\mathbf y})]\,
({\Phi}^{-1}){}^{(y)}_{\;\;\;\;(u)}
({\mathbf y},{\mathbf u})=
\,{\delta}^3({\mathbf y}-{\mathbf u})\,.
\]
(The boundary conditions of this operator depend on a concrete choice
of a base manifold $M$.)
Notice that in the formula (\ref{5}), the matrix ${\Phi}^{-1}$,  
as well as the other terms of
the projector  $N$, is given on the gauge surface $ \Sigma$.

In our principal bundle, the 
  metric $d_{(x, y)}$ on the orbit over the base point is determined by using the Killing vectors $K_{( y)}$:
\[
d_{(x,y)}=K^{(i,z)}_{\;\;\;\;\;\;(x)}
G_{(i,z)(j,u)}K^{(j,u)}_{\;\;\;\;\;\;( y)}+K^{(a,z)}_{\;\;\;\;( x)}G_{(a,z)(b,u)}K^{(b,u)}_{\;\;\;\;(y)} \]
That is,
\[
 d_{(x, y)}=\Bigl[-\triangle (\mathbf x)+
G_{ab}(\bar J_{})^a_c (\bar J_{})^b_{c'} f^c(\mathbf x)f^{c'}(\mathbf x)
\Bigr]\delta ^3(\mathbf x-\mathbf y)
\]
\[
 =\gamma_{(x,y)}(\mathbf x,\mathbf y)+\gamma'_{(x,y)}(\mathbf x,\mathbf y)
\]
($d_{(x, y)}=[-\triangle (\mathbf x)+g_0^2\vec f^2(\mathbf x)]\delta ^3(\mathbf x-\mathbf y)\equiv\triangle_f\delta ^3(\mathbf x-\mathbf y)$.)

An ``inverse matrix'' to the ``matrix'' $d_{(x)(y)}$ is defined by the following equation:
\[
 d_{(x,y)}d^{(y,z)}=\delta^{( z)}_{(x)}=
\delta ^3(\mathbf z-\mathbf x).
\]
In explicit  form this equation is written as follows:
\begin{eqnarray*}
&&\Bigl[-\triangle (\mathbf x)+G_{ab}(\bar J_{})^a_c (\bar J_{})^b_{c'}{\tilde f}^c(\mathbf x){\tilde f}^{c'}(\mathbf x)
\Bigr]d^{(x)(z)}
\nonumber\\
&&\;\;\;\;\;\;\;\;\;\;\;\;\;\;\;\;\;\;\;\;=\delta ^3(\mathbf z-\mathbf x).
\end{eqnarray*}
Thus, $d^{(y,z)}$ is the Green function of the operator given by the expression in  square brackets. It is assumed  that a certain boundary condition for the equation is chosen.

In the principal fiber bundle ${\mathrm P}(\mathscr P\times_{\tilde\mathscr G}\mathscr V,\mathscr G)$ there is a connection one-form known as the ``Coulomb connection'' (or ``mechanical connection''). It is defined as follows:
\[
 \hat\omega =\Bigl(\mathscr A^{(x)}_{\;\;\;\;(j, y)}\,d A^{\ast (j, y)}+\mathscr A^{(x)}_{\;\;\;(a, y)}d{\tilde f}^{(a,y)}\Bigr)+da^{}(\mathbf x),
\]
where the components of the connection are given by
\[
 \mathscr A^{(x)}_{\;\;\;\;(j, y)}=d^{(x,z)}K^{( k, v)}_{\;\;\;\;\;\;(z)}G_{(k,v)(j,y)}=[\partial_j(y)d^{(x,y)}]
\]
and
\[
 \mathscr A^{(x)}_{\;\;\;(p, y)}=d^{(x,z)}K^{(a, v)}_{\;\;\;\;\;\;(z)}G_{(a,v)(p,y)}=d^{(x,y)}(\bar J_{})^a_c\tilde f^c(\mathbf y)G_{ap}.
\]

In the new coordinate basis $(\frac{\delta}{\delta A^{\ast}{}^{(i,x)}},\frac{\delta}{\delta {\tilde f}^{(b,x)}},\frac{\delta}{\delta a^{(x)}})$
the original metric  of the  manifold $\mathscr P \times \mathscr V$ can be rewritten as follows:
\begin{equation}
\displaystyle
{\tilde G}_{\cal A\cal B}(A{}^{\ast},\tilde f,a)=
\left(
\begin{array}{ccc}
 \tilde G_{(i,x)(j,y)} & 0 & \tilde G_{(i,x)(z)}\\
 0 & \tilde G_{(a,x)(b,y)} & \tilde G_{(a,x),(z)}\\
\tilde G_{(j,y)(u)} & \tilde G_{(b,x)(u)} & d_{(u,z)}\\
\end{array}
\right),
\label{metric2c}
\end{equation}
where 
\begin{eqnarray*}
{\tilde G}_{(i,x)\,( j,y)}&=&
G_{({m},\tilde x)\,(n,\tilde y)}
\left(P_{\bot}\right)^{({m},\tilde x)}_
{\;\;\;\;\;\;(i,x)}\;
\left(P_{\bot}\right)^{(n,\tilde y)}_
{\;\;\;\;\;\;(j,y)}\;\nonumber\\
&&=\delta_{i n}
\left(P_{\bot}\right)^{(n,x)}_{\;\;\;\;\;\;(j,y)}.
\nonumber\\
\end{eqnarray*}
\begin{eqnarray*}
&&{\tilde G}_{(i,x)\,(y)}=
G_{(m,u)\,( n,v)}
\left(P_{\bot}\right)^{( m,u}_
{\;\;\;\;\;\;(i,x)}\;
K^{(n,v)}_{\;\;\;(y)}\,\nonumber\\
&&\;\;\;\;\;\;\;\;\;\;\;\;\;\;\;\;\;
={\delta}_{mn}\,\int d\bar v
\left(P_{\bot}\right)^{( m,v)}_
{\;\;\;\;(i,x)}\,[\partial^n(\bar v)\delta^3(\bar v-\bar y)].
\end{eqnarray*}
${\tilde G}_{(x)\,(y)}$=$d_{(x)\,(y)}$, 
${\tilde G}_{(a,x)\,(b,y)}=\delta_{ab}\delta^3(\mathbf x-\mathbf y)$,
${\tilde G}_{(a,x)\,(z)}=\delta_{ab}\tilde J^b_c\tilde f^c(x)\delta^3(\mathbf x-\mathbf z)$.

In the finite-dimensional case, the projection operator onto the ``gauge surface'' $\Sigma $ is given by the following expression: 
\[
(P_{\bot})^A_B={\delta}^A_B-{\chi}^{\mu}_B\;({\chi}
{\chi}^{\top})^{-1}{}^{\nu}_{\mu}\;({\chi}^{\top})^A_{\nu}\,.
\]
In gauge theories it can be viewed as a symbolic expression of the corresponding projection operator. Using appropriate substitutions, one can show that for the Coulomb gauge in our case of scalar electrodynamics
\[
{\chi}^{(\nu, x)}_{\;\;\;\;(i,y)}=
\left[\,{\partial }_i(\mathbf x)\;{\delta}^3 (\mathbf x-\mathbf y)\,\right]\,.
\]
  this operator is given by the expression
  \begin{eqnarray*}
&&\bigl(P_{\bot}\bigr)^{(k,x)}_{\;\;\;\;\;\;
(m,y)}=\nonumber\\
&&\left({\delta}^k_{\,m}\;
{\delta}^3(\mathbf y-\mathbf x)
+ {\partial}_m(\mathbf y)\int d^3 u\, K(\mathbf y,\mathbf u)\,\left({\partial}^k(\mathbf u)
\,{\delta}^3(\mathbf u-\mathbf x)\right)\;\right)
\end{eqnarray*}
Thus, the projection operator  can be written symbolically as
\[
\bigl(P_{\bot}\bigr)^{(k,x)}_{\;\;\;\;\;\;(m,y)}=
\left[{\delta}^k_{\,m}+
{\partial}_m\frac{1}{(-{\partial}^2)}
{\partial}^k\right]
{\delta}^3(\mathbf y-\mathbf x)\,.
\]
The pseudoinverse matrix ${\tilde G}^{\mathcal A,\mathcal B}(A^*, \tilde f, a)$ to  matrix (\ref{metric2c}) can be derived from the corresponding matrix which was obtaned in finite-dimensional case
\begin{equation}
\displaystyle
\left(
\begin{array}{ccc}
{G}^{EF}N_E^AN_F^B & G^{EF}N^A_EN^a_F &  G^{EF}N^A_E{\Lambda}^{\alpha}_F\\
G^{EF}N^A_FN^b_E & G^{ab}+G^{EF}N^a_EN^b_F & G^{EF}N^b_E{\Lambda}^{\alpha}_F\\
G^{EF}N^B_F{\Lambda}^{\beta}_E & G^{EF}N^a_E{\Lambda}^{\alpha}_F &G^{EF}{\Lambda}^{\beta}_E{\Lambda}^{\alpha}_F  \\
\end{array}
\right),
\label{metric2b}
\end{equation}
where as projection operators $N$ we should take the operators that we used earlier. And instead of  ${\Lambda}^{\nu}_E\equiv({\Phi}^{-1})^{\nu}_{\mu}(Q{}^{\ast}){\chi}^{\mu}_E(Q{}^{\ast})$ one 
should insert $  \Lambda^{(z)}_{(i,u)}$ which is given by
\[
 \Lambda^{(z)}_{(i,u)}=\int d\mathbf v (\Phi^{-1})^{(z)}_{(v)}(\mathbf z-\mathbf v)[\partial_i(\mathbf  v)\delta^3(\mathbf v-\mathbf u)]. 
\]

The pseudoinversion of ${\tilde G}_{\cal A\cal B}$ means that
\[
\displaystyle
{\tilde G}^{\tilde{\cal A}\tilde{\cal D}}{\tilde G}_{\tilde{\cal D}\tilde{\cal B}}=
\left(
\begin{array}{ccc}
  (P_{\bot})^A_B & 0 & 0\\
 0 & {\delta}^a_b & 0\\
0 & 0 & {\delta}^{\alpha}_{\beta}\\
\end{array}
\right).
\]

\section{Stochastic differential equations in reduced space}
 The next step after introducing the coordinates of the principal fiber bundle on the manifold $\tilde\mathscr P$ should be the transformation of the original stochastic differential equations. This is important because the measure in the path integral we used was determined by the solutions of these equations. It is known how this can be done in the finite-dimensional case. In \cite{Storchak_int_model} the path integral reduction was studied for a model finite-dimensional mechanical system, which is in many ways analogous to the gauge system with interaction considered here. We could repeat the same thing described in this article in relation to our gauge system. Instead, we will try to directly generalize the results obtained there by appropriately rewriting them in terms of gauge fields. Of course, this is based on the assumption that everything done there can be extended to gauge theories.

In our work, the main point after introducing adapted coordinates in the obtained stochastic equations was that the drift term was rewritten so that it corresponded to the standard term when describing diffusion on the submanifold. It turned out that in addition to the standard form, consisting of terms with the Christoffel coefficients and the mean curvature of the orbit space, there is another term associated with the mean curvature of the orbit. The next step was to transform the path integral, which led to a factorization of the measure in the integral. This was based on the use of stochastic differential equations from the theory of random processes. And the  last step was the Girsanov transformation by which the additional drift term became the ``quantum'' correction to the interaction potential in the Hamilton operator. Therefore, our task is to rewrite for fields the results obtained in the finite-dimensional case at each stage.

After introducing the coordinates associated with the fiber bundle,  we obtain in \cite{Storchak_int_model} the following stochastic differential equations: 
\begin{eqnarray*}
&&dQ^{\ast}{}^A(t)=\mu^2\kappa\Bigl(-\frac12h^{\tilde B \tilde M}\,{}^{ \mathrm  H}{\tilde \Gamma}^{ A}_{\tilde B\tilde M}+j^{ A}_{\Roman 1}+j^{ A}_{\Roman 2}\Bigr)dt
 +\mu\sqrt{\kappa}N^A_C\mathscr X^C_{\bar M}dw^{\bar M}_t,
 \label{sde_Q_ast_j}
 \\
  &&d\tilde f^a(t)=\mu^2\kappa\Bigl(-\frac12h^{\tilde B \tilde M}\,{}^{ \mathrm  H}{\tilde \Gamma}^{ a}_{\tilde B\tilde M}+j^{ a}_{\Roman 1}+j^{ a}_{\Roman 2}\Bigr)dt
  +\mu\sqrt{\kappa}\bigl(N^a_C\mathscr X^C_{\bar M}dw^{\bar M}_t+\mathscr X^a_{\bar b}dw^{\bar b}_t\bigr).
  \label{sde_f_j}
 \end{eqnarray*}
 In these equations, $h^{\tilde B\tilde M}=G^{EF}N^{\tilde B}_EN^{\tilde M}_F$ (the tilde sign above the script means that the sum applies to all indexes it has, both uppercase and lowercase),
 \begin{eqnarray*}
 j^{ A}_{\Roman 1}=\frac12h^{BM}N^A_{B, M}+\frac12h^{\tilde B \tilde M}\Bigl(\, {}^{ \mathrm  H}{\tilde \Gamma}^{ A}_{\tilde B\tilde M}-N^A_C\,{}^{ \mathrm  H}{\tilde \Gamma}^{ C}_{\tilde B\tilde M}\Bigr),
\nonumber\\
\end{eqnarray*}
\begin{eqnarray*}
 j^{ a}_{\Roman 1}=-\frac12N^{a}_{C}\Bigl(\,h^{BM}N^C_{B, M}+h^{\tilde B\tilde M}\,{}^{ \mathrm  H}{\tilde \Gamma}^{ C}_{\tilde B\tilde M}\Bigr)
 =\frac12h^{CM}N^a_{C,M}-\frac12N^a_Ch^{\tilde B\tilde M}\,{}^{ \mathrm  H}{\tilde \Gamma}^{ C}_{\tilde B\tilde M}.
\label{j_a_deriv_N}
 \end{eqnarray*}
We do not write out the equation for the group variable, since it has a standard form and is not important for what follows. In the resulting equations we see two additional types of terms $j_{\Roman 1}$ and $j_{\Roman 2}$, which are associated with the mean curvature of the orbit space and the mean curvature of the orbit, respectively. To describe stochastic evolution on a submanifold, the equation requires terms with the mean curvature of the orbit space and with Christoffel coefficients. Therefore, only these terms remain in the final equations. First, we consider how the terms with the Christoffel coefficients are represented in our field model. Then the same will be done with the terms $j_{\Roman 1}$ and $j_{\Roman 2}$, using their explicit expression, which is also given.

The Christoffel symbols 
${}^{ \mathrm  H}{\tilde \Gamma}^{\tilde R}_{BM}$ were determined (modulo such terms
 $T^{\tilde M}_{BC}$ that satisfy  ${\tilde G}^{\mathrm H}_{A\tilde M}T^{\tilde M}_{BC}=0$) using the following equality: 
\[
 {}^{ \mathrm  H}{\tilde \Gamma}_{BM\tilde D}={\tilde G}^{\mathrm H}_{\tilde R\tilde D}{}^{ \mathrm  H}{\tilde \Gamma}^{\tilde R}_{BM}.
\]
This is also true for other symbols, such as ${}^{ \mathrm  H}{\tilde \Gamma}_{BMa}={\tilde G}^{\mathrm H}_{\tilde Ra}{}^{ \mathrm  H}{\tilde \Gamma}^{\tilde R}_{BM}$ etc. And ${}^{ \mathrm  H}{\tilde \Gamma}_{BM\tilde D} $  calculated for the horizontal metric
\begin{eqnarray*}
\left(
\begin{array}{cc}
\tilde G^{\rm H}_{AB} & \tilde G^{\rm H}_{A a}\\
\tilde G^{\rm H}_{b B} & \tilde G^{\rm H}_{ab}\\
\end{array}
\right),
\end{eqnarray*}
where
${\tilde G}^{\rm H}_{AB}=G_{AB}-G_{AC}K^C_{\mu}d^{\mu\nu}K^D_{\nu}G_{DB},$
$\;\;\tilde G^{\rm H}_{Aa}=-G_{AB}K^B_{\mu}d^{\mu\nu}K^b_{\nu}G_{ba}$, $\;\tilde G^{\rm H}_{ba}=G_{ba}-G_{bc}K^c_{\mu}d^{\mu\nu}K_{\nu}^pG_{pa}$.

Similar Christoffel coefficients were calculated earlier in \cite{Storchak_lagr_Poinc_YM} for a Yang-Mills field interacting with a scalar field. By reducing them to our simpler case, we can extract the desired result. Such a constraint, represented symbolically (as before using the representation for the finite-dimensional case), looks like this:

\begin{center}
\bf{Christoffel symbols for the  horizontal metric in scalar electrodynamics}
\end{center}
\begin{eqnarray*}
 { }^{\;\;\rm H}{\Gamma}_{BM}^A=
 G^{AR}{ }^{\;\;\rm H}{\Gamma}_{BMR}&=&0
{}\\
{ }^{\;\;\rm H}{\Gamma}_{Bm}^A=
G^{AR}{ }^{\;\;\rm H}{\Gamma}_{BmR}&=&-\frac12\mathscr A^{\beta}_{B,m}K^A_{\beta}
\\
{}^{\;\;\rm H}{\Gamma}_{pq}^A=
G^{AR}{ }^{\;\;\rm H}{\Gamma}_{pqR}&=&-\frac12\bigl(\mathscr A^{\beta}_{p,q}K^A_{\beta}+\mathscr A^{\beta}_{q,p}K^A_{\beta})
{}\\
{ }^{\;\;\rm H}{\Gamma}_{mB}^A=
G^{AR}{ }^{\;\;\rm H}{\Gamma}_{mBR}&=&-\frac12\mathscr A^{\beta}_{m,B}K^A_{\beta}
{}\\
{ }^{\;\;\rm H}{\Gamma}_{AB}^r=
G^{rm}{ }^{\;\;\rm H}{\Gamma}_{ABm}&=&
 +\frac12(K^r_{\mu, p}K^p_{\sigma})(\mathscr A^{\sigma}_{A}\mathscr A^{\mu}_{B}+\mathscr A^{\mu}_{A}\mathscr A^{\sigma}_{B}).
{}\\
{ }^{\;\;\rm H}{\Gamma}_{pB}^r=
G^{rm}{ }^{\;\;\rm H}{\Gamma}_{pBm}&=&-\frac12\mathscr A^{\beta}_{B,p}K^r_{\beta}-\mathscr A^{\beta}_{B}K^r_{\beta,p}\\
&{}& +\frac12(K^r_{\varepsilon, q}K^q_{\mu})(\mathscr A^{\mu}_{p}\mathscr A^{\varepsilon}_{B}+\mathscr A^{\mu}_{B}\mathscr A^{\varepsilon}_{p}).
{}\\
{ }^{\;\;\rm H}{\Gamma}_{pq}^r=
G^{rm}{ }^{\;\;\rm H}{\Gamma}_{pqm}&=&-\frac12\bigl(\mathscr A^{\beta}_{p,q}K^r_{\beta}+\mathscr A^{\beta}_{q,p}K^r_{\beta})
-(\mathscr A^{\beta}_{p}K^r_{\beta,q}+\mathscr A^{\beta}_{q}K^r_{\beta,p})\\
&{}& +\frac12(K^r_{\mu, n}K^n_{\nu})(\mathscr A^{\mu}_{q}\mathscr A^{\nu}_{p}+\mathscr A^{\nu}_{q}\mathscr A^{\mu}_{p}).
\end{eqnarray*}
These Christoffel symbol representations can be easily transformed into field theory representations.

As for the expressions for the mean curvatures of the orbit space, in scalar electrodynamics the terms on the right-hand side of their definitions (terms with derivatives) disappear. The remaining parts use only Christoffel symbols in their definitions.

Now consider the expression for the mean curvature of the orbits. The components of this vector are as follows:

\begin{eqnarray*}
&& j_{\Roman 2}^A=-\frac12 \,G^{CC'}N^A_{C'}N^{B'}_CG_{BB'}\,d^{\alpha \beta}\bigl(\nabla_{K_{\alpha}}K_{\beta}\bigr)^B,
\nonumber\\
&&j_{\Roman 2}^a=-\frac12d^{\alpha \beta}\Bigl(N^a_B\,\bigl(\nabla_{K_{\alpha}}K_{\beta}\bigr)^B+\bigl(\nabla_{K_{\alpha}}K_{\beta}\bigr)^a\Bigr).
\label{j_2}
\end{eqnarray*}
These expression can be rewritten using  the following identities:
\begin{eqnarray*}
 &&d^{\alpha \beta}\bigl(\nabla_{K_{\alpha}}K_{\beta}\bigr)^B=-\frac12\Bigl(G^{BC}N^A_C{\sigma}_A+G^{BC}N^a_C\,{\sigma}_a\Bigr),
\nonumber\\
&&d^{\alpha \beta}\bigl(\nabla_{K_{\alpha}}K_{\beta}\bigr)^a=-\frac12 G^{aq}{\sigma}_q,
\end{eqnarray*}
where ${\sigma}_A=d^{\alpha \beta}\frac{\partial }{\partial Q^{\ast A}}d_{\alpha \beta}\equiv \frac{\partial }{\partial Q^{\ast A}}\ln d$ and ${\sigma}_a=\frac{\partial }{\partial {\tilde f}^a} \ln d,$   $d=\det d_{\alpha \beta}$.

As a result, we obtain 
\begin{eqnarray*}
\displaystyle
\Bigl(
\begin{array}{cc} 
j_{\Roman 2}^A\\
j_{\Roman 2}^a
\end{array}
\Bigr)
&=&\frac14 
\Bigl(
\begin{array}{cc}
h^{AC} & h^{Ab}\\
h^{aC}& h^{ab}\\
\end{array}\Bigr)
\Bigl(
\begin{array}{cc}
 {\sigma}_C\\ 
{\sigma}_b\\
\end{array}
\Bigr)
\label{j_2_sigma}
\end{eqnarray*}

The following path integral transformation, associated with the Girsanov transformation of the stochastic process, leads to a reduction Jacobian whose exponential part is given by
\begin{equation}
J=-\frac18\mu^2\kappa\bigl(\triangle_{\tilde{\Sigma}}^{{\scriptscriptstyle \rm H}}\sigma +\frac14<\partial\sigma,\partial \sigma>_{\tilde{\Sigma}}\bigr),
\label{jacobian_reduct}
\end{equation}
where 
$$\triangle_{\tilde{\Sigma}}^{{\scriptscriptstyle \rm H}}\sigma=
h^{AB}\sigma_{AB}+2h^{Ab}\sigma_{Ab}
+h^{ab}\sigma_{ab}
-h^{\tilde B \tilde M}\,{}^{ \mathrm  H}{\tilde \Gamma}^{ A}_{\tilde B\tilde M}\sigma_A
-h^{\tilde B \tilde M}\,{}^{ \mathrm  H}{\tilde \Gamma}^{ a}_{\tilde B\tilde M}\sigma_a$$
and 
\[
 <\partial\sigma,\partial \sigma>_{\tilde{\Sigma}}=h^{AB}{\sigma}_A {\sigma}_B+2h^{aB}{\sigma}_a {\sigma}_B+h^{ab}{\sigma}_a {\sigma}_b.
\]
For a finite-dimensional system used in scalar electrodynamics, these expressions can be simplified by neglecting derivatives with respect to variables with capital letters. This follows from the fact that $d_{\alpha\beta}$ corresponds to the metric defined on orbits in scalar electrodynamics. There it is set as
$d_{(x, y)}=[-\triangle (\mathbf x)+g_0^2\vec f^2(\mathbf x)]\delta ^3(\mathbf x-\mathbf y)\equiv\triangle_f\delta ^3(\mathbf x-\mathbf y)$, which is independent of the electromagnetic field. Thus contribution on $J$ is only from the scalar  field. 
That is, we will have 

$$\tilde\triangle_{\tilde{\Sigma}}^{{\scriptscriptstyle \rm H}}\sigma=+h^{ab}\sigma_{ab}
-h^{\tilde B \tilde M}\,{}^{ \mathrm  H}{\tilde \Gamma}^{ a}_{\tilde B\tilde M}\sigma_a$$
and 
\[
 <\partial\sigma,\partial \sigma>_{\tilde{\Sigma}}'=h^{ab}{\sigma}_a {\sigma}_b.
\]
Now  the forward Kolmogorov equation at $\kappa =i$ becomes
the Schr\"odinger equation with the Hamilton operator 
$\hat H=-\frac{\hbar}{\kappa}{\hat H}_{\kappa}\bigl|_{\kappa =i}$
\[
\hat{H}_{\kappa}=
\frac{\hbar \kappa}{2m}\tilde\triangle _{\tilde{\Sigma}}-\frac{\hbar \kappa}{8m}\Bigl[\tilde\triangle_{\tilde{\Sigma}}^{{\scriptscriptstyle \rm H}}\sigma +\frac14<\partial\sigma,\partial \sigma>'_{\tilde{\Sigma}}\Bigr]+\frac{1}{\hbar \kappa}\tilde V.
\]
Note also that the geometric properties of the operator $\tilde\triangle _{\tilde{\Sigma}}$,
$$\tilde\triangle_{\tilde{\Sigma}} =\tilde\triangle_{\tilde{\Sigma}}^{{\scriptscriptstyle \rm H}}+2j^{ A}_{\Roman 1}\,\partial_A+2j^{ a}_{\Roman 1}\,\partial_a.$$

In conclusion we note that the problem we encounter in scalar electrodynamics when we perform the quantum reduction procedure is the calculation of the determinant of the differential operator. It is possible that even before the procedure is started, it is necessary to regularize the quantities that eventually lead to infinities, as was done in \cite{Paycha,Arn_Paycha,Arn_Paycha_2}. But in this case, it will be necessary to change, apparently, the method of reduction that works well in the finite-dimensional case. The main thing, however, as is evident from the examples of reduction using path integrals in quantum mechanics, is that one cannot neglect the consideration of the terms arising related to the volume of the orbit. This was also confirmed in the quantization of pure Yang-Mills fields in Lott's work\cite{Lott}, where the zeta function method was used to regularize a similar term.

\end{document}